\documentclass[prb,aps, twocolumn,floatfix,showpacs,showkeys, superscriptaddress ]{revtex4-1}

\usepackage{graphicx}
\usepackage{epstopdf}

\usepackage{color}

\usepackage[utf8]{inputenc}
\usepackage{xcolor}

\begin{document}

\title{Charge density wave and large non-saturating magnetoresistance \\in YNiC$_2$ and LuNiC$_2$}

\author{Kamil K. Kolincio}
\email{kamkolin@pg.edu.pl}
\affiliation{Faculty of Applied Physics and Mathematics, Gdansk University of Technology,
Narutowicza 11/12, 80-233 Gdansk, Poland}
\affiliation{RIKEN Center for Emergent Matter Science (CEMS), Wako, Saitama 351-0198, Japan}

\author{Marta Roman}
\affiliation{Faculty of Applied Physics and Mathematics, Gdansk University of Technology,
Narutowicza 11/12, 80-233 Gdansk, Poland}

\author{Tomasz Klimczuk}
\affiliation{Faculty of Applied Physics and Mathematics, Gdansk University of Technology,
Narutowicza 11/12, 80-233 Gdansk, Poland}

\begin{abstract}
We report a study of physical properties of two quasi-low dimensional metals YNiC$_2$ and LuNiC$_2$ including the investigation of transport, magnetotransport, galvanomagnetic and specific heat properties. In YNiC$_2$ we reveal two subsequent transitions associated with the formation of weakly coupled charge density wave at $T_{CDW}$ = 318 K, and its locking in with the lattice at $T_1$  = 275 K. These characteristic temperatures follow the previously proposed linear scaling with the unit cell volume, demonstrating its validity extended beyond the lanthanide-based $R$NiC$_2$. We also find that, in the absence of magnetic ordering able to interrupt the development of charge density wave, the Fermi surface nesting leads to opening of small pockets, containing high mobility carriers. This effect gives rise to substantial enhancement of magnetoresistance, reaching 470 \% for YNiC$_2$ and 50 \% for LuNiC$_2$ at $T$ = 1.9 K and $B$ = 9 T.
\end{abstract}

\maketitle

\section{Introduction}
The large variety of unique physical phenomena offered by quasi low-dimensional systems arouse the unfading interest of researchers exploring the field of condensed matter physics. Large anisotropy of the electronic structure is a precursor for the Peierls transition towards charge density wave (CDW) with electronic carrier condensation and Fermi surface nesting\cite{Monceau2012, Gruner2000, Gruner1988}.
Low dimensionality has been suggested to play a crucial role in high temperature superconductivity while charge density wave has been found to emerge in all the phase diagrams of the cuprate superconductors\cite{Caprara2017, Caplan2017, Laliberte2018, Cyr2018}. This fact additionally amplifies the interest in the interactions between various types of electronic, quantum, and magnetic ordering\cite{Jacques2014, Chang2012, Chang2016, Graf2004, Gruner2017, Cho2018}. 

Recently, extensive attention has been devoted to the family of ternary carbides $R$NiC$_2$, where $R$ is rare earth element.  This group of materials offers the rare opportunity to tune the magnetic ground state via replacement of the rare earth atom (the nickel atoms carry no magnetic moment). Most of the lanthanide-based members of this family are antiferromagnets ($R$ = Ce, Nd, Gd, Tb, Dy, Ho, Er and Tm) with N\'eel temperature varying from 3.4 K for HoNiC$_2$ to 25 K for TbNiC$_2$\cite{Schafer1997, Onodera1995, Onodera1998, Hanasaki2011, Roman2018_1, Steiner2018}; SmNiC$_2$ is a ferromagnet with $T_C$ = 17.5 K \cite{Onodera1998}, PrNiC$_2$ shows a weak magnetic anomaly at $T^*$ = 8 K, LaNiC$_2$ is a noncentrosymmetric superconductor with $T_{c}$ = 2.9 K\cite{Wiendlocha2016, lee_superconductivity_1996, Pecharsky1998, Quintanilla2010, Landaeta2017} while LuNiC$_2$ was recently reported as a plain paramagnet down to 1.9 K\cite{Steiner2018}. Furthermore, most of the $R$NiC$_2$ compounds (with the exception of $R$ = La and Ce) show charge density wave with Peierls  temperature ($T_{CDW}$) not only higher than $T_N$, $T_C$ and $T^*$, but exceeding 300 K for the late lanthanides ($R$ = Ho - Lu)\cite{Murase2004, Ahmad2015, Michor2017, Laverock2009, Shimomura2016}. So far, the charge density wave has never been found in $R$NiC$_2$ compounds, with $R$ outside of the lanthanide group. The electronic structure calculations however, have revealed the resemblance between the Fermi surface topology of YNiC$_2$\cite{Hase2009} and lanthanide-based $R$NiC$_2$\cite{Laverock2009, Steiner2018} showing CDW. Alas, the fermiology of this compound has not been discussed in the terms of nesting.

Remarkably, magnetic ordering has been found to mutually interact with the CDW; in SmNiC$_2$ the ferromagnetic transition destroys the charge density wave\cite{Shimomura2009, Hanasaki2012, Kim2012}, while in NdNiC$_2$ and GdNiC$_2$ antiferromagnetism only partially suppresses the CDW, and both entities coexist below $T_N$\cite{Yamamoto2013, Kolincio20161, Hanasaki2017, Kolincio2017}. On the other hand, this magnetic anomaly has been found to enhance the nesting properties in PrNiC$_2$ \cite{Yamamoto2013, Kolincio2017} and some signatures of a constructive influence of CDW on AFM were recently observed in GdNiC$_2$ \cite{Kolincio20161, Hanasaki2017}, Nd$_{1-x}$Gd$_x$NiC$_2$\cite{Roman2018_2}, and Nd$_{1-x}$La$_x$NiC$_2$ \cite{Roman_tmp} despite the clear competition between these types of ordering.
For the majority of the the $R$NiC$_2$ family members, the CDW state is influenced (mostly suppressed) by magnetism, and therefore the Fermi surface nesting is partially or even completely disturbed. 

In this paper we study the physical properties of paramagnetic YNiC$_2$ and LuNiC$_2$ in order to explore the consequences of a pure and fully developed Peierls transition in $R$NiC$_2$ in the absence of magnetic ordering. We report for the first time the charge density wave in YNiC$_2$ with $T_{CDW}$ = 318 K followed by a putative lock-in transition at $T_1$ = 275 K. We also observe the large positive and remarkably linear magnetoresistance, which at the lowest temperatures reaches 470 \% for YNiC$_2$ and 50 \% for LuNiC$_2$ with no signs of saturation up to $B$ = 9 T. By detailed analysis of magnetotransport and galvanomagnetic properties we find that this effect stems from the multiband character of the electrical conductivity and existence of the high mobility pockets remaining in the Fermi surface after uninterrupted, yet imperfect nesting characteristic for quasi-2D metals undergoing a Peierls transition\cite{Monceau2012, Gruner2000, Gruner1988}.

\section{Experimental}
The polycrystalline samples of YNiC$_2$ and LuNiC$_2$  were synthesized by arc melting of elemental precursors (no excess of Y or Lu was added) followed by annealing at 900$^o$ C, according to the procedure described in detail in ref. \cite{Kolincio2017}. Powder x-ray diffraction (pXRD) was conducted with PANalytical X’Pert PRO-MPD diffractometer using K$_\alpha$ line of Cu spectrum. Experimental data was analyzed with the use of Fullprof software\cite{FULLPROF}.

All the physical properties measurements were performed with a commercial Quantum Design Physical Properties Measurement System (PPMS) in the temperature range from 1.9 to 400 K and magnetic field up to 9 T. Thin ($\phi=37 \mu m$) Pt wires serving as electrical contacts for four-probe transport and Hall measurements were spark-welded to the polished samples surfaces. The magnetic field for Hall and magnetoresistance measurements was oriented perpendicularly to the current direction.
The Hall signal was measured with the reversal of the direction of the magnetic field and corrected for the parasitic longitudinal resistance component via antisymmetrization of the measured data. Specific heat measurements were conducted with a standard relaxation method on flat samples with polished surfaces. Apiezon N grease was used as a heat conducting medium for measurements in temperature range 1.9 K - 300 K. Since Apiezon N reveals a glass transition above 300 K, \cite{Bunting1969, Kreitman1971}, the data collection at elevated temperatures was performed with Apiezon L.

\section{Results and discussion}
\subsection{X-ray diffraction}
\begin{figure} [t]
  \includegraphics[angle=0,width=1.0\columnwidth]{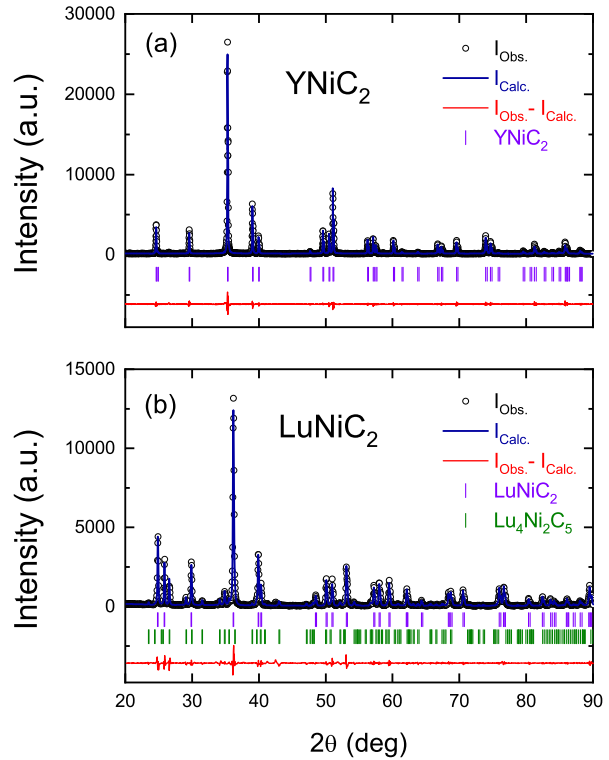}
 \caption{\label{XRD} Powder x-ray diffraction patterns (black open points) with LeBail fit refinement (blue solid line) for YNiC$_2$ (a) and LuNiC$_2$ (b). Bragg peak positions for YNiC$_2$ and LuNiC$_2$ phases are marked by vertical violet lines. Similar, yet green lines in panel (b) apply to the impurity phase Lu$_4$Ni$_2$C$_5$. Difference between observed and calculated pattern is represented by solid red line.}
  \end{figure}

The quality and phase purity of polycrystalline YNiC$_2$ and LuNiC$_2$ samples were confirmed with pXRD. The analysis of the obtained diffraction patterns, shown in Fig. \ref{XRD} revealed that the observed peaks for both compounds are successfully indexed in the orthorombic CeNiC$_2$-type structure with space group Amm2 (\#38). Only for LuNiC$_2$, additional reflections corresponding to Lu$_4$Ni$_2$C$_5$ impurity phase were detected. From the comparison of the highest peaks corresponding to main and impurity phases respectively, we have estimated the relative amount of Lu$_4$Ni$_2$C$_5$ to 9\%. The pXRD results for both $R$NiC$_2$ compounds were analyzed by LeBail refinement which revealed lattice constants of YNiC$_2$: $a$ = 3.5733(1) \AA, $b$ = 4.5082(1) \AA, $c$ = 6.0351(1) \AA~ and of LuNiC$_2$: $a$ = 3.4468(1) \AA, $b$ = 4.4734(2) \AA, $c$ = 5.9787(2) \AA. These structural parameters are in good agreement with previous reports\cite{Jeitschko1986, Steiner2018}.

\subsection{Transport and magnetoresistance}
\begin{figure*} [ht!]
  \includegraphics[angle=0,width=2.1\columnwidth]{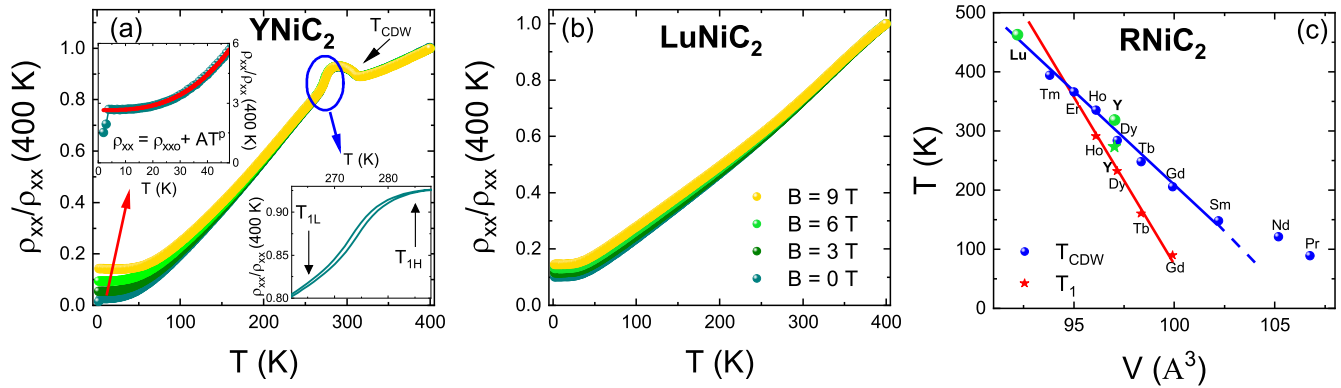}
 \caption{\label{RT} (a), (b) Temperature dependence of the normalized electrical resistivity $\rho_{xx}/\rho_{xx}(400 K)$(T) measured in magnetic fields varying from 0 to 9 T for YNiC$_2$ (a) and LuNiC$_2$ (b) respectively. Upper inset of panel (a) depicits the resistivity of YNiC$_2$ in the low temperature limit. Fit to the data above superconducting transition with the equation \ref{EQres} is shown with red solid line. Lower inset of shows the thermal hysteresis limited from below and above with temperatures $T_{1L}$ and  $T_{1H}$ respectively, in the vicinity of the lock-in transition. Legend for pannels (a) and (b) is displayed on panel (b). (c) the characteristic temperatures $T_{CDW}$ and $T_1$ for YNiC$_2$ and LuNiC$_2$ compared with the analogous values for the other members of the $R$NiC$_2$ family. The solid lines correspond to the linear scaling with unit cell volume ($V$), reported  previously \cite{Roman2018_1}. Green points correspond to the YNiC$_2$ and LuNiC$_2$ compounds studied in this paper. The Peierls temperature for LuNiC$_2$ has been determined in ref. \cite{Roman2018_1, Steiner2018}, while to the authors knowledge, there are no previous reports on CDW in YNiC$_2$ }
  \end{figure*}
  
The thermal dependence of the resistivity of YNiC$_2$ and LuNiC$_2$ is depicted respectively in panels a and b of Fig. \ref{RT}. At high temperatures, the transport properties of YNiC$_2$ show a regular metallic behavior with the resistivity lowering as the temperature is decreased. At $T$ = 318 K a local minimum followed by a hump is observed in $\rho_{xx}$($T$). At lower temperatures, the resistivity returns to the metallic character with $
\frac{d\rho_{xx}}{dT}<0$. Such behavior is typical of a Peierls transition in quasi-2D metals, where the nesting is not complete and a certain number of free carriers remain in the Fermi surface. One unusual feature here, however, is the sharp decrease of the resistivity at temperatures slightly below the maximum of $\rho_{xx}$($T$) indicated by an arrow in the Fig. \ref{RT}a. This stands in contrast to a smooth crossover observed for canonical Peierls transitions\cite{McMillan1977}. To determine the character of this additional anomaly, we have performed slow heating and cooling temperature sweeps in the vicinity of the CDW transition (0.1 K/min). The results, shown in the lower inset of Fig. \ref{RT}a reveal the presence of a narrow thermal hysteresis opening approximately at  $T_{1H}\simeq$  285 K and closing at  $T_{1L}\simeq$ 265 K. The transition temperature is determined as the average of  $T_{1H}$ and  $T_{1L}$, giving $T_1$ = 275 K. Such a first order transition is expected for the preexisting incommensurate CDW locking in with the lattice and transforming into a commensurate modulation\cite{McMillan1975} as reported beforehand for the late lanthanide-based $R$NiC$_2$\cite{Shimomura2016, Roman2018_1}. Interestingly both the Peierls and presumed lock-in transition temperatures stand in agreement with the previously proposed linear scaling with the unit cell volume\cite{Roman2018_1}, as shown in Fig. \ref{RT}c. One must, however, remember that despite the analogy with the other $R$NiC$_2$, an x-ray diffuse scattering experiment performed with single crystals is required to deliver an ultimate evidence of the lock-in nature of the transition seen at $T_1$ = 275 K. An additional feature observed in the low temperature resistivity curve is a decrease of $\rho_{xx}$ at $T$ = 4 K (upper inset of Fig. \ref{RT}a), reminiscent of the onset of a superconducting transition. The absence of any signature of the bulk superconductivity in specific heat at this temperature allows us to attribute this anomaly to the presence of a trace amount of YC$_2$ impurity\cite{Gulden_1997}, not detected by x-ray diffraction. To estimate the genuine value of $\rho_{xx}$ for YNiC$_2$ at lowest $T$, we have fitted the temperature range 5 K $<T<$ 50 K with the power law:
\begin{equation}
\label{EQres}
\rho_{xx} = \rho_{xx0}+AT^p,
\end{equation}
(where $\rho_{xx0}$ is the residual resistivity and exponent $p$ depends on the prevailing scattering mechanism) and extended the obtained function to lowest temperature.

 For LuNiC$_2$ (see Fig. \ref{RT}c) the character of the conductivity is metallic up to 400 K, in agreement with previous reports of a similar metal-metal CDW transition at $T_{CDW}\simeq$ 450 K\cite{Roman2018_1, Steiner2018}, which is beyond the upper temperature limit of the PPMS, thus not revealed by our current measurement. Interestingly, despite that the CDW modulation wavevector determined previously for LuNiC$_2$ \cite{Steiner2018} differs from the vectors reported for $R$ = Pr, Nd, Sm, Gd and Tb \cite{Yamamoto2013, Shimomura2009, Shimomura2016}, the Peierls temperature corresponding to Lu bearing compound (reported in references\cite{Roman2018_1, Steiner2018}) obeys the scaling proposed for magnetic $R$NiC$_2$ \cite{Roman2018_1} as well, which is shown in Fig. \ref{RT}c.

\begin{figure*} [ht!]
  \includegraphics[angle=0,width=2.1\columnwidth]{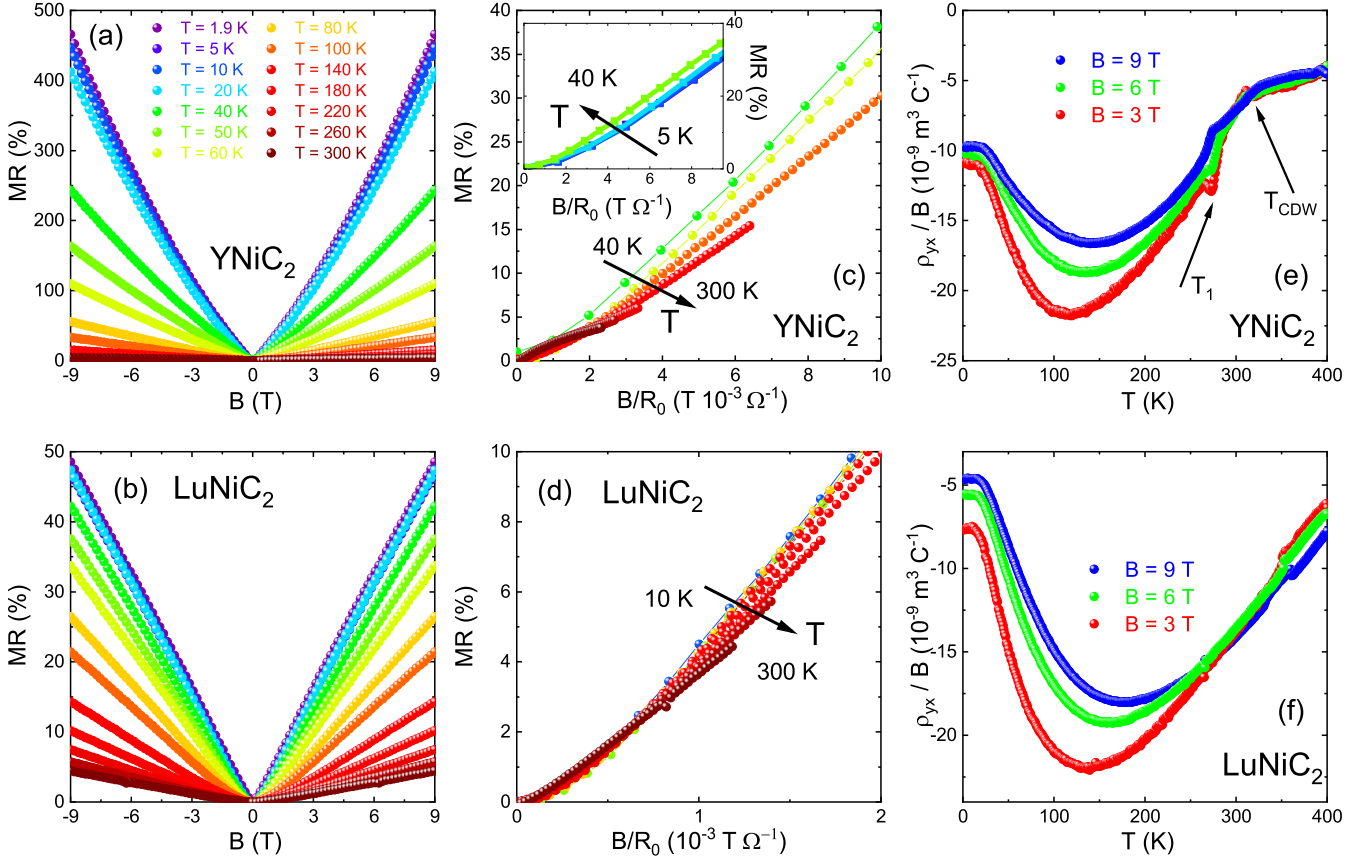}
 \caption{\label{Panel1} (a)-(b) Magnetoresistance in YNiC$_2$ (a) and LuNiC$_2$ (b) as a function of applied magnetic field. (c)-(d) Kohler plots of the magnetoresistance in YNiC$_2$ (c) and LuNiC$_2$ (d) Inset of panel (c) shows the low temperature range for YNiC$_2$ (e)-(f) Thermal dependence of the normalized Hall resistivity $\frac{\rho_{yx}}{B}$ in YNiC$_2$ (e) and LuNiC$_2$ (f). The legend for (a), (b), (c) and (d) is displayed on panel (a).}
\end{figure*}

  An external magnetic field has a negligible impact on the transport properties in the normal state of YNiC$_2$, typical for a casual metal. On the other hand, the application of $B$ significantly increases the electrical resistivity in the CDW state of this material. This is also true, however, to a lesser extent, for the CDW state of LuNiC$_2$.
The magnetic field dependence of magnetoresistance defined as:
\begin{equation}
\label{EQMR}
MR=\frac{\rho_{xx}(B)-\rho_{xx}(B=0)}{\rho_{xx}(B=0)}\cdot 100\%,
\end{equation} 
is presented in Fig. \ref{Panel1}a and b for YNiC$_2$ and LuNiC$_2$, respectively. For YNiC$_2$, the MR curve for $T$ = 1.9 K where the zero field value of $\rho_{xx}$ was estimated from equation \ref{EQres} is close to the one measured for $T$ = 5 K, above the superconducting transition and where the $\rho_{xx}$ is already close to the residual value. This result confirms that, the MR($B$) curve for $T$ = 1.9 K is not contaminated by any spurious contribution that could stem from the superconducting impurity phase.
At low field, the electrical resistance of both compounds follows a MR$\sim B^q$ dependence on magnetic field. MR is subquadratic ($q \simeq$  1.4 at $T$ = 1.9 K) for YNiC$_2$ and approximately parabolic ($q \simeq$ 2.2 K at $T$ = 1.9 K ) for LuNiC$_2$. As the magnetic field is raised, the character of MR$(B)$ curves for both compounds evolves towards a linear manner, without any signs of saturation up to 9 T. The magnitude of the magnetoresistance increases as the temperature is lowered 
and at $T$ = 1.9 K and $B$ = 9 T reaches 470 \% for the former and 50 \% for the latter compound, respectively. 
Typically, the linear magnetoresistance term is described by one of two prevailing models: the classical approach of Parish and Littlewood based on mobility fluctuations in inhomogeneous material\cite{Parish2003, Parish2005} or Abrikosov's quantum model used beyond the quantum critical limit, when only a single Landau level is occupied\cite{Abrikosov1969, Abrikosov1988, Abrikosov2000}. In a CDW metal however, the linear MR can also originate from the fluctuations of the charge density wave order parameter enhancing the scattering in certain regions of the Fermi surface \cite{Sinchenko2017, Frolov2018} or from sharp curves on the carrier path due to nesting induced reduction of FS \cite{Feng_2019}. Another possible scenario is that the linearity in MR originates from the complicated geometry of the Fermi surface (which is additionally modified by nesting), containing both closed and open orbits, which in a polycrystalline sample contribute to MR with saturating and non-saturating signals \cite{Lifshitz1957} respectively.

It shall be noted that, such a large positive magnetoresistive effect has never been reported for any members of the $R$NiC$_2$ family. Typically, in $R$NiC$_2$ compounds exhibiting Peierls transition, MR shows a negative sign due to partial or complete suppression of charge density wave, induced either by magnetic field or magnetic ordering\cite{Yamamoto2013, Hanasaki2012, Kolincio20161, Kolincio2017, Hanasaki2017, Lei2017}. In  YNiC$_2$ and LuNiC$_2$ however, no signatures of magnetism has been found above $T$ = 1.9 K.

The magnetoresistance can be discussed in the framework of the semiclassical Kohler approach\cite{Pippard2009}. The prediction of this model is that on the condition of uniform scattering over the whole Fermi surface and a single type of electronic carriers with constant concentration, all the plots:
\begin{equation}
\label{EQKohler}
MR=f(\omega_c\tau)=f\left( \frac{B}{\rho_{xx}(B=0)}\right)
\end{equation}
(where $\tau$ and $\omega_c$ are relaxation time and cyclotron frequency, respectively) superimpose into a single line. The plot of MR as a function of 
$\frac{B}{R_0}$ ($R_0$ is the zero field resistance) for YNiC$_2$ is shown in Fig. \ref{Panel1}c. The deviations from Kohler scaling, shown in Fig. \ref{Panel1}c, are weak yet visible. Upon decreasing the temperature, the plots are subsequently moved higher on the vertical scale (faster increase of MR($\frac{B}{R_0}$)) which is observed in a wide temperature range above $T$ = 40 K. For $T<$ 40 K, curves are pressed slightly lower (slower $\frac{B}{R_0}$ growth), as depicted in the inset of Fig. \ref{Panel1}c. At lowest temperatures, where the zero field resistance is already close to the residual value, the curves superimpose (within the experimental resolution).
  As shown in Fig. \ref{Panel1}d, for LuNiC$_2$ the divergence from Kohler's scaling is even less pronounced but still visible. The strongest effect is seen in the temperature range $T>\frac{T_{CDW}}{2}$, where the CDW gap is expected not to be completely open yet\cite{Gruner1988}. This violation of the Kohler rule can be attributed both to the reconstruction of the Fermi surface due to nesting and to presence of more than one type of carriers in the CDW state\cite{McKenzie1998, Yasuzuka2005}. We suggest that a stronger manifestation of the deviation from the MR scaling could be observed at temperatures in the close vicinity of $T_{CDW}$ as in tungsten bronzes also showing Peierls transition \cite{Kolincio20162}. This range is, however, beyond the scope of our experimental equipment.

\subsection{Hall effect}
\begin{figure*} [ht!]
  \includegraphics[angle=0,width=2.1\columnwidth]{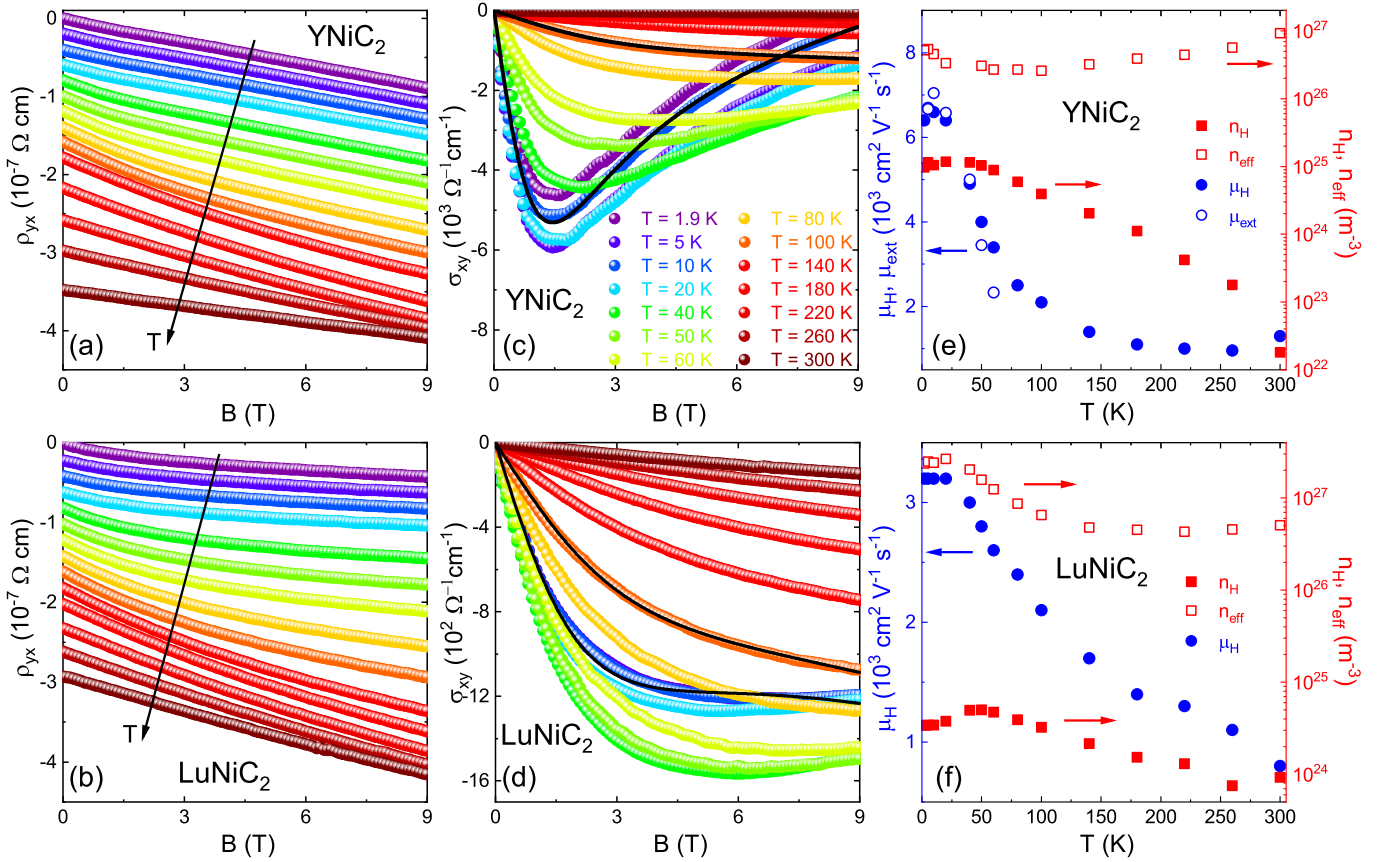}
 \caption{\label{Panel2} (a)-(b) Magnetic field dependence of Hall resistivity $\rho_{yx}$ in YNiC$_2$ (a) and LuNiC$_2$(b). The plots have been vertically shifted for clarity and the vertical scale applies to the plot for corresponding to $T$ = 1.9 K. (c)-(d) Hall conductivity $\sigma_{xy}$ in YNiC$_2$ (c) and LuNiC$_2$ (d). The black solid lines are representative fits to the experimental data with equation \ref{EQsigmaxyAP}. (e)-(f) The results of the analysis of Hall resistivity and conductivity: mobilities $\mu_H$, $\mu_{ext}$ and concentrations $n_H$, $n_{eff}$ plotted as a function of temperature for YNiC$_2$ (e) and LuNiC$_2$ (f). The legend for (a), (b), (c) and (d) is displayed in panel (c).}
\end{figure*}

To explore the evolution of carrier concentrations, we have examined the Hall effect for both compounds. The thermal dependence of the Hall resistivity ($\rho_{yx}$) is depicted in Fig. \ref{Panel1}e (YNiC$_2$) and \ref{Panel1}f (LuNiC$_2$). For YNiC$_2$, $\frac{\rho_{yx}}{B}$ is almost temperature independent above $T_{CDW}$. At this characteristic temperature, the Hall resistivity shows an abrupt downturn, indicating the loss of free electrons due to the CDW condensation. The presumed lock-in transition is indicated by a kink in  $\frac{\rho_{yx}}{B}(T)$. At lower temperatures, the Hall resistance shows a minimum and then returns to less negative values. Previously, such an effect was observed in magnetic $R$NiC$_2$, and was attributed both to the suppression of charge density wave by the magnetic ordering and to the onset of the anomalous component of the Hall effect\cite{Kim2012, Kolincio20161, Kolincio2017, Lei2017}. Due to the absence of long range magnetism in YNiC$_2$, these two terms appear to be irrelevant in this case.
At temperatures below $T_1$, the  $\frac{\rho_{yx}}{B}(T)$ curves do not superimpose into a single line which suggests that in the CDW state,  $\rho_{yx}$ is not linear with $B$. 

For LuNiC$_2$ the Peierls temperature $T_{CDW}\simeq$ 450 K\cite{Roman2018_1, Steiner2018}, thus at 400 K, which is the maximum temperature limit of our experiment, the system is already in the charge density wave state. All the curves reveal a kink at $T\simeq$ 355 K. Its origin is not clear, however, while this weak anomaly is not detected by other measurements, it might result from the experimental artifact instead of being truly intrinsic to the sample. Another scenario is, that this anomaly originates from the Lu$_4$Ni$_2$C$_5$ impurity phase. Similarly to YNiC$_2$, the sign of $\rho_{yx}$ is negative in the whole temperature range, indicating the dominance of electrons. This is not the only similarity between the $\frac{\rho_{yx}}{B}$ curves for both compounds. Here we also find that for LuNiC$_2$ the Hall resistivity is also driven to more negative values as the free electrons are condensed in the CDW state, which is followed by the return of $\rho_{yx}$ to close to zero at lower temperatures.  We find that the $\frac{\rho_{yx}}{B}$ superimpose at temperatures above approximately 250 K. At lower temperatures, the plots do not coincide with each other, indicating a nonlinearity of $\rho_{yx}(B)$ also in LuNiC$_2$. Similarly to the case of YNiC$_2$, further temperature decrease leads to the upturn of the Hall resistivity, which also cannot be attributed to magnetic ordering. A plausible scenario to explain these features is the existence of more than one type of electronic carrier, originating from unnested pockets remaining in the Fermi surface after imperfect nesting, a situation characteristic of quasi-2D metals showing charge density wave\cite{Monceau2012, Gruner2000}. 

To obtain a more detailed picture of the electronic parameters, we have examined the magnetic field dependence of $\rho_{yx}$. The results of field sweeps at constant temperatures, shown in the Fig. \ref{Panel2}a for YNiC$_2$ and \ref{Panel2}b for LuNiC$_2$ reveal a visible deviation of Hall signal from linearity.
In the absence of long range magnetic interactions or ordering, this effect is a clear manifestation of the multiband character of electrical conductivity\cite{Akiba2017, Li2016, Liu2017, Luo2015, Wang2014}. 
In the two-band model, the Hall resistivity is expressed with equation (\ref{EQHall})\cite{Hurd1972}:
\begin{equation}
\label{EQHall}
\frac{\rho_{yx}}{B}=\frac{1}{e}\frac{n_h\mu_h^2-n_e\mu_e^2+(n_h-n_e)\mu_e^2\mu_h^2B^2}{(n_h\mu_h+n_e\mu_e)^2+(n_h-n_e)^2\mu_h^2\mu_e^2B^2}
\end{equation}
where $n_h$, $n_e$, $\mu_h$ and $\mu_e$ are respectively concentrations and mobilities corresponding to two (hole and electron) conduction channels. The direct $\rho_{yx}$ fit with eq. (\ref{EQHall}) gives four dependent parameters, which may lead to misguiding conclusions\cite{Rotella20151}. However, the high field  limit of this equation gives an approximate measure of the effective carrier concentration $n_{eff}$\cite{Sun2014}, which will be discussed in section D:
\begin{equation}
\label{EQHallHF}
\frac{\rho_{yx}}{B}=\frac{1}{e}\frac{1}{n_h-n_e}=\frac{1}{e}\frac{1}{n_{eff}}
\end{equation}
\subsection{Multiband conductivity}
More detailed information can be extracted by transforming components of resistivity tensor  $\rho_{yx}$ and $\rho_{xx}$ to obtain  Hall conductivity $\sigma_{xy}$ via the following equation:

\begin{equation}
\label{EQsigmaxy}
\sigma_{xy}(B)=\frac{\rho_{yx}}{\rho_{yx}^2+\rho_{xx}^2}
\end{equation}

In the multiband system, $\sigma_{xy}$ is a superposition of the terms originating from subsequent contributing bands.  Equation (\ref{EQsigmaxy}) can be then rewritten as\cite{Lin2016}:

\begin{equation}
\label{EQsigmaxymulti}
\sigma_{xy}(B)=\sum_{i}\frac{\sigma_i\mu_iB}{1+\mu_i^2B^2}
\end{equation}

Hall conductivity is commonly used to determine the electronic parameters, since the extremum of $\sigma_{xy}(B)$ is a direct measure (or at least a good approximation in a multiband system) of the dominant mobility $\mu_{ext}$ calculated from the inverse of the magnetic field $B_{ext}$, at which $\sigma_{xy}$ peaks\cite{Liang2014}:

\begin{equation}
\label{EQmuinv}
\mu_{ext}=\frac{1}{B_{ext}}
\end{equation}

The Hall conductivity for both compounds is negative in the whole temperature range and at low temperatures shows a minimum, which for YNiC$_2$ is visibly sharper than for LuNiC$_2$. The position of this minimum shifts from high fields to lower values of $B$ as temperature is lowered. For YNiC$_2$, the {$B_{ext}$ is clearly defined, while the broad extremum seen in LuNiC$_2$ precludes the precise determination of the peak position. Since the direct fitting of $\sigma_{xy}$ with equation \ref{EQsigmaxymulti} assuming one hole and one electron bands once again requires using four dependent parameters, for further analysis we have used an approach\cite{Takahashi2011, Ishiwata2013}, in which we have assumed the existence of a single band with high mobility carriers and the remaining band(s) to show significantly lower mobility:

\begin{equation}
\label{EQsigmaxyAP}
\sigma_{xy}(B)=n_{xy}e\mu_{xy}^2B  \left( \frac{1}{1+\mu_{xy}^2B^2} +C_{xy} \right)
\end{equation}

Equation \ref{EQsigmaxyAP} allows the estimation of the mobility $\mu_{xy}$, and concentration $n_{xy}$ of this single 'fast' band (pocket), while other 'slower' bands contribute to $C_{xy}$ parameter. The typical fits are shown by solid lines in panels c and d of Fig. \ref{Panel2} respectively. We have found that $\sigma_{xy}$ can be reasonably well-described with equation (\ref{EQsigmaxyAP}) despite of the fact that the zero field values of $\rho_{xx}$ can be significantly increased due to the polycrystalline character of the samples. 

\begin{figure} [ht!]
  \includegraphics[angle=0,width=1.1\columnwidth]{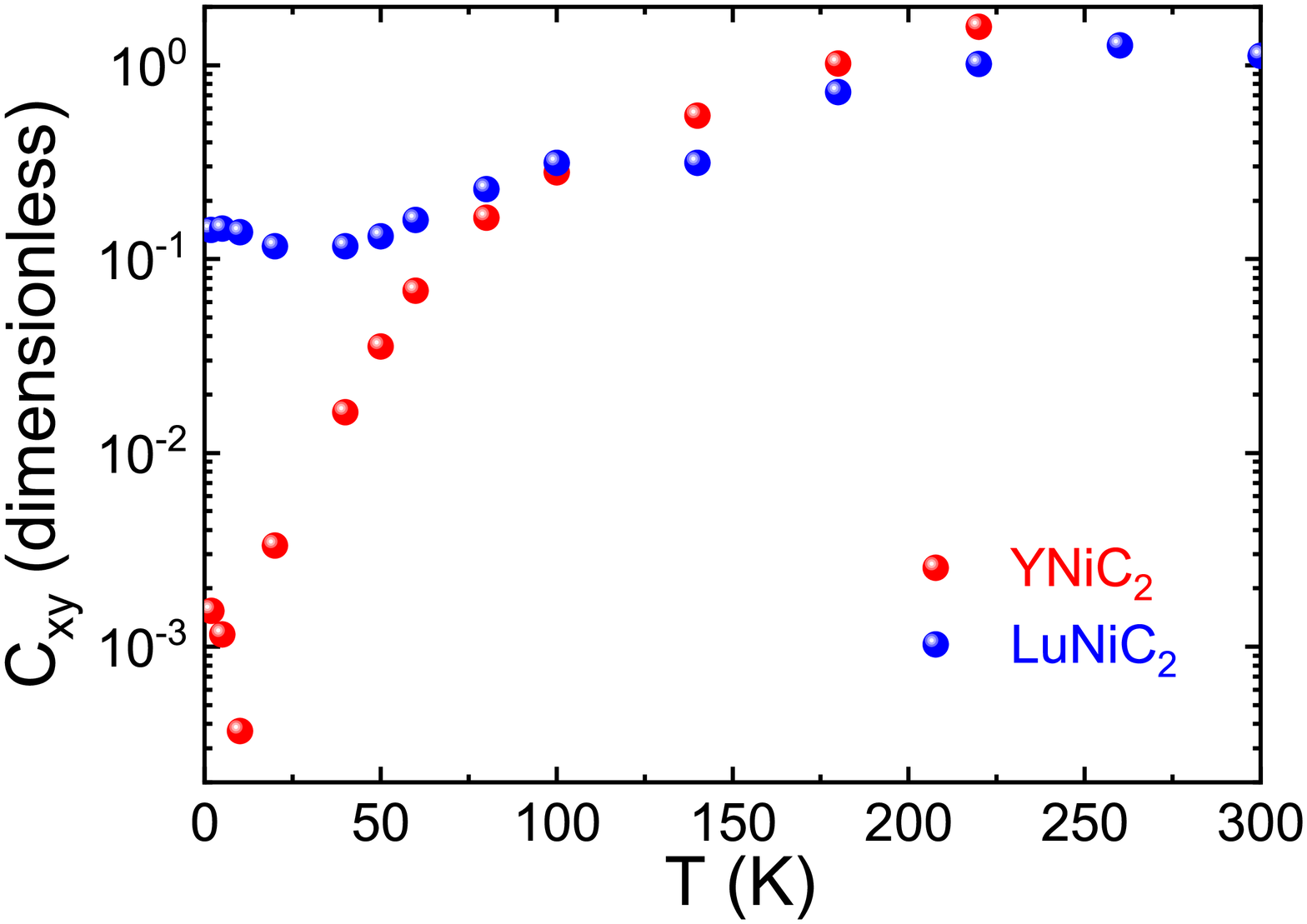}
 \caption{\label{Cpar} $C_{xy}$  parameters resulting from least square fit of $\sigma_{xx}$ with equation \ref{EQsigmaxyAP} for YNiC$_2$ (red color) and LuNiC$_2$ (blue color).}
  \end{figure}

The parameters derived from this procedure, as well as the values of $n_{eff}$ and $\mu_{ext}$, are summarized in Fig.  \ref{Panel2}e for YNiC$_2$ and \ref{Panel2}f, for LuNiC$_2$. The mobilities $\mu_{ext}$ and $\mu_{xy}$ coincide with each other for YNiC$_2$, and both quantities reach very large values of $7\cdot 10^3$ cm$^2$ V$^{-1}$ s$^{-1}$ at $T$ = 1.9 K. 
The electronic mobility $\mu_{xy}$ of LuNiC$_2$ is twice as  small as in the case YNiC$_2$, yet still considerable.  The coincidence of $\mu_{xy}$ and $\mu_{ext}$ is an additional argument for the correctness of the value calculated from $\sigma_{xy}$. It shall be, however, noted that, next to the increase of the residual resistivity, the polycrystalline samples character is expected also to substantially lower the electronic mobility in comparison with the single crystal.

 As seen in Fig. \ref{Panel2}e and f, for both compounds, the concentration of the carriers originating from the high mobility band increases as temperature is lowered. 
 The growth of $n_{xy}$  is concomitant with the decrease of the effective carrier concentration $n_{eff}$ below Peierls temperature. This is consistent with the nesting picture: while the majority of electrons are removed from the conducting band and condenses towards CDW, the parallel opening of unnested pockets results in the increase of the high mobility carriers. Interestingly, while the results of the $\sigma_{xy}$ analysis suggests the electron origin of the carriers described with concentration $n_{xy}$, the upturn of Hall resistivity and $n_{eff}$ at lowest temperatures can possibly be caused by the existence of holes with not so large mobility, thus contributing only to $C_{xy}$ parameter in equation (\ref{EQsigmaxyAP}). The temperature interval in which this effect is observed corresponds to the range in which the a turnover of deviations from Kohler scaling is observed in YNiC$_2$ (inset of Fig. \ref{Panel1}c).
 
The $C_{xy}$ parameter serves as an estimate of the ratio of the conductivities stemming from 'slow' to 'fast' bands respectively. Thermal dependence of $C_{xy}$ for both compounds is shown in figure \ref{Cpar}. For both compounds, the values of this parameter show values close to unity at high $T$ and  decrease as temperature is lowered.  $C_{xy}$ reaches $\simeq$ 0.001 for YNiC$_2$ and $\simeq$ 0.1 for LuNiC$_2$. Small upturn is seen at low temperatures at low temperatures, which can be associated with the existence of an additional band as suggested above. The relatively low values of $C_{xy}$, especially in the former compound, underline the major role played by the carriers originating from the 'fast' pocket in the terms of electronic transport and show that the used approximate model can be used to describe the properties of YNiC$_2$ and LuNiC$_2$. 

The presence of both electron and hole pockets in the CDW state of LuNiC$_2$ is also consistent with the results of band structure calculations\cite{Steiner2018}. Owing to the similarities between the Fermi surfaces of YNiC$_2$ \cite{Hase2009} and other $R$NiC$_2$ showing CDW, it is reasonable to assume the relevance of the same scenario for Y bearing compound as well. 

The high mobility of carriers contained in these pockets is then likely responsible for the high magnitude of MR in both compounds. Opening of such pockets was reported in a number of quasi-2D CDW materials showing strong, yet imperfect Fermi surface nesting, leading to the enhancement of magnetoresistance\cite{Rotger1994, Rotger1996, Yasuzuka1999, Chen2017},
 themopower\cite{Rhyee2009, Rhyee2015} and galvanothermomagnetic properties\cite{Bel2003, Kolincio20163}.

This result supports the scenario of strong Fermi surface reconstruction in YNiC$_2$ and LuNiC$_2$, which is possible due to the absence of any competing magnetic ordering which was responsible for the CDW suppression in majority of $R$NiC$_2$ family members \cite{Yamamoto2013, Hanasaki2012, Kolincio20161, Kolincio2017, Hanasaki2017, Lei2017}.

\subsection{Specific heat}

To complement the results of transport, magnetotransport, and Hall experiments, and to further characterize the CDW transition in YNiC$_2$, we have measured the specific heat $C_p$. Fig. \ref{CP}a depicts the temperature dependence of the specific heat capacity $C_p(T)$ in the temperature range 1.9 - 300 K. At 300 K, $C_p$ reaches approximately 80\% of the value expected by Dulong-Petit law (3nR $\sim$ 100 J mol$^{-1}$ K$^{-1}$), suggesting that the Debye temperature for YNiC$_2$ exceeds 300 K. 

No anomalies have been detected at low temperatures, which confirms the absence of bulk superconductivity or magnetic ordering. 
The specific heat data plotted as $\frac{C_p}{T} $ vs. $T^2$ presented in Fig. \ref{CP}b has been fitted to the equation (\ref{CPeq}) with both sides divided by $T$.

\begin{equation}
\label{CPeq}
C_p  =  \gamma T + \beta T^3
\end{equation}

\noindent where the first and second terms represent electronic and lattice contributions, respectively.
The fit revealed values of Sommerfeld coefficient $\gamma$ = 1.65(1) mJ mol$^{-1}$K$^{-2}$ and $\beta$ = 0.326(4) mJ mol$^{-1}$K$^{-4}$, the latter corresponds to the Debye temperature $\Theta_D$ = 620 K according to:

\begin{equation}
\label{Debye}
\Theta_D = \left(  \frac{12\pi^4 nR}{5\beta} \right)^{\frac{1}{3}} 
\end{equation}

\noindent where R =  8.314 J mol$^{-1}$K$^{-1}$  and $n$ is the number of atoms per formula unit ($n$ =  4 for YNiC$_2$). This value is larger than the $\Theta_D$ = 456 K reported previously for YNiC$_2$ \cite{long_heat_2001}. The Debye temperature found here is also larger than the value reported for LaNiC$_2$ ($\Theta_D$ = 445 K) \cite{Prathiba2016}. Such behavior can be reasonably explained by a mass relationship: for molar mass of Y smaller than La, one expects higher $\Theta_D$.

  \begin{figure} [t]
  \includegraphics[angle=0,width=1.0\columnwidth]{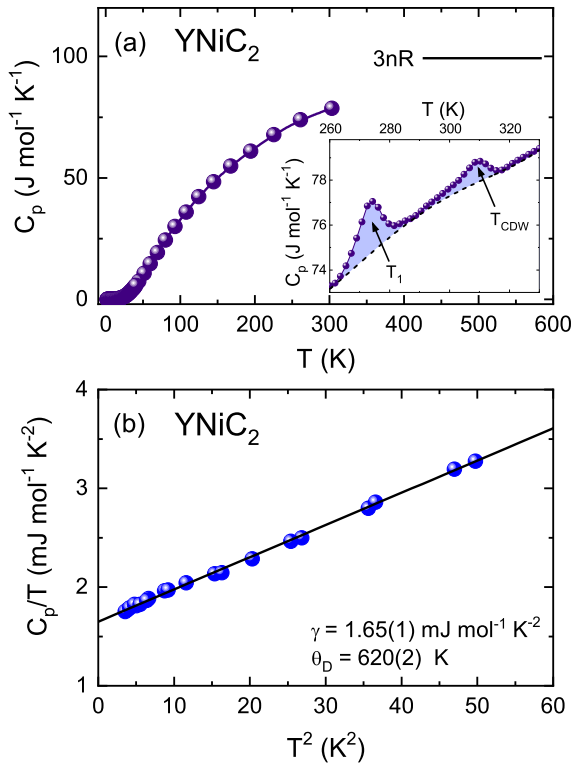}
 \caption{\label{CP} (a) Specific heat of YNiC$_2$ as a function of temperature. The inset shows an expanded view on the vicinity of the Peierls transition. The anomalies are marked with arrows. Dashed line corresponds to the background subtracted to evaluate the excess specific heat corresponding to the transitions - highlighted with light violet color. The high temperature measurements were performed with Apiezon L grease - see experimental section for details. (b) $\frac{C_p}{T} (T^2)$ in the low temperature region. Black solid line corresponds to the fit with equation (\ref{CPeq}), divided by $T$ on both sides.}
  \end{figure}
  
Results of the detailed measurements of $C_p(T)$ above room temperature are shown in the inset of Fig. \ref{CP}a. The Peierls transition is signaled by a small maximum of $C_p(T)$ at $T$ = 310 K, being in rough agreement with the transition temperature $T_{CDW}$ established from resistivity measurements. The relative increase of specific heat at the charge density wave formation temperature denotes $\frac{\Delta C_p}{C_p(T_{CDW})} \simeq 1.1 \% $, thus is at the same order of magnitude as in canonical CDW systems as NbSe$_3$\cite{Tomic1981}, K$_{0.9}$Mo$_6$O$_{17}$\cite{Escribe1984}, or tungsten bronzes\cite{Chung1993}. 

The mean-field weak coupling description of the Peierls transition predicts the specific heat jump of:

\begin{equation}
\label{BCSeq}
\frac{\Delta C_p}{\gamma T_{CDW}}=1.43
\end{equation}

In the case of YNiC$_2$, the equation (\ref{BCSeq}) gives the value of 1.79, slightly larger than the BCS prediction, indicating the relevance of a weak coupling scenario.

Visibly stronger and sharper anomaly accompanies the presumed lock-in crossover at $T_1$ = 275 K. Here the specific heat increases by $\frac{\Delta C_p}{C_p(T_1)} \simeq 2.9 \% $, with $C_p(T_1)$ estimated from the background. The magnitude of this anomaly is noticeably larger than for the features typically observed at the incommensurate-commensurate CDW transformation\cite{Craven1977, Kuo2004}.

The entropy $\Delta S$ and enthalpy $\Delta H$ of both anomalies were estimated from the excess specific heat at each transition by integrating the $\frac{\Delta C_p}{T}dT$ of and $\Delta C_pdT$ respectively, after evaluating and subtracting the background values of $C_p$.
The integrated regions are highlighted by light violet color in Fig. \ref{CP}a.
\begin{table}[t!]
\caption{Thermodynamic parameters: relative increase of specific heat $\frac{\Delta C_p}{C_p(T)}$, entropy $\Delta S$ and enthalpy $\Delta H$ at transition temperatures $T_{CDW}$ and $T_1$ in YNiC$_2$.}
\label{tableHC}
\begin{ruledtabular}
\begin{tabular}{cccc}
& $\frac{\Delta C_p}{C_p(T)}$ (\%) & $\Delta S$ (mJ mol$^{-1}$K$^{-1}$) & $\Delta H$ (J mol$^{-1}$)\\
 \hline
$T_{CDW}$ & 1.1 & 30.6 & 9.4\\
$T_1$ & 2.9 & 77.8 & 21.3\\
\end{tabular}
\end{ruledtabular}

\end{table}
The results of the integration of $C_p$ excess  accompanying the phase transitions are summarized in Tab. \ref{tableHC}. While the size of $\Delta C_p$ step at $T_{CDW}$ stands in agreement with the BCS predictions as well as with the values found in other materials exhibiting a weakly coupled charge density wave, we find an unusualy low value of $\Delta S$ accompanying this transition. This can be imposed by the high Peierls temperature, resulting in a large denominator of $\frac{\Delta C_p}{T}$ and thus small result of the integral. The value of enthalpy however, does not diverge from the typically observed values in CDW systems\cite{Escribe1984, Wang2006}.
In agreement with the comparison of $\Delta C_p$ jump, for the crossover at $T_1$, the values $\Delta S$ and $\Delta H$, are significantly larger than for the Peierls transition at $T_{CDW}$. This result is unexpected, since typically the lock-in transition is not associated with the opening of a new electronic gap, next the one already existing in the CDW state.  
The sharp peak shape of this anomaly can suggest a large role played by CDW order parameter fluctuations \cite{McMillan1977, Kuo2001, Kwok1990}. The detailed analysis of crystal structure, as well as of the phonon spectra, performed on a single crystal is required to elucidate this issue.

\section{conclusions}
We have examined the physical properties of polycrystalline YNiC$_2$ and LuNiC$_2$. The former compound shows at $T_{CDW}$ = 318 K Peierls transition with signatures of BCS - mean field weak coupling scenario, followed by presumed lock-in crossover at $T_1$ = 275 K. The  temperatures corresponding to these anomalies, revealed by transport, Hall effect and specific heat measurements, are found to obey the linear scaling with the unit cell volume, observed previously with lanthanide-based $R$NiC$_2$ compounds. 
Both studied materials show large magnetoresistance in the CDW state, reaching 470 \% for YNiC$_2$ and 50 \% for LuNiC$_2$ at $T$ = 1.9 K and $B$ = 9 T. To discuss its origin, we have combined the analysis of thermal and magnetic field depencence of Hall effect and magnetoresistance. We have found that the effect standing behind such strong magnetoresistive features in YNiC$_2$ and LuNiC$_2$ is the existence of pockets, including at least one with high mobility carriers, remaining in the Fermi surface after nesting, caused by fully developed CDW transition not interrupted by competing orders such as magnetism or superconductivity.

\begin{acknowledgments}
 The authors gratefully acknowledge the financial support from National Science Centre (Poland), grant number:  UMO-2015/19/B/ST3/03127. The authors would also like to thank to M. Hirshberger and C. Zhang (both RIKEN), Alain Pautrat (CRISMAT), Helen Walker (ISIS), Nathan Runyon, and Jesse Sprowes for their helpful advice. 
 \end{acknowledgments}
 
%

\end{document}